# Towards culturally-appropriate conversational AI for health in the majority world: An exploratory study with citizens and professionals in Latin America

**Dorian Peters**, Imperial College London and University of Cambridge, UK, d.peters@imperial.ac.uk

**Fernanda Espinoza**, Imperial College London and La Victoria Lab, Peru, fernanda@lavictoria.pe

**Marco Da Re**, Imperial College London, UK, m.dare@imperial.ac.uk

**Guido Ivetta**, Universidad Nacional de Córdoba, Argentina, guidoivetta@mi.unc.edu.ar

**Luciana Benotti**, Universidad Nacional de Córdoba, Argentina, luciana.benotti@unc.edu.ar

**Rafael A. Calvo**, Imperial College London, UK, r.calvo@imperial.ac.uk

## Abstract

There is justifiable interest in leveraging conversational AI (CAI) for health across the majority world, but to be effective, CAI must respond appropriately within culturally and linguistically diverse contexts. Therefore, we need ways to address the fact that current LLMs exclude many lived experiences globally. Various advances are underway which focus on top-down approaches and increasing training data. In this paper, we aim to complement these with a bottom-up locally-grounded approach based on qualitative data collected during participatory workshops in Latin America. Our goal is to construct a rich and human-centred understanding of: a) potential areas of cultural misalignment in digital health; b) regional perspectives on chatbots for health and c) strategies for creating culturally-appropriate CAI; with a focus on the understudied Latin American context. Our findings show that academic boundaries on notions of culture lose meaning at the ground level and technologies will need to engage with a broader framework; one that encapsulates the way economics, politics, geography and local logistics are entangled in cultural experience. To this end, we introduce a framework for 'Pluriversal Conversational AI for Health' which allows for the possibility that more relationality and tolerance, rather than just more data, may be called for.

## 1 Introduction

Both healthcare access and health conversation are shaped by culture. Conversational AI (CAI) directed at the health domain will, therefore, need to respond appropriately within diverse cultural contexts. This will be particularly challenging for the many regions globally that are under-represented or not represented in training data. In this paper, we focus on the particular cultural context of Latin America, with hopes our findings can reveal insights that also contribute to other regions in the majority world. Our goal is to contribute to the collective understanding of what health chatbots will need to account for within a culturally and linguistically diverse context.

Cultural biases and misalignment in Large Language Models (LLMs) have been widely documented (Atari et al., 2023; Frankfurt, 1971; Gallegos et al., 2024; Pawar et al., 2024). These issues are also increasingly addressed at Natural Language Processing conferences (e.g. Prabhakaran et al., 2024) which gather interdisciplinary teams to identify and tackle the technical challenges presented by cultural misalignment. For example, there is increasing work comparing LLMs with responses by humans to Global Value Surveys (Cao et al., 2023; C. Li et al., 2024; Tao et al., 2024). But these quantitative approaches lack the depth of qualitative research and community involvement in envisioning potential culturally-situated uses of NLP and AI. As the review by Liu et al.contends (2024) "Getting the process right requires consultation with target communities".

To contribute to addressing this, we conducted eight workshops with Latin Americans from different cultural backgrounds in Lima (Peru), Huancayo (Peru), Cordoba (Argentina), Carlos Paz (Argentina) and with mixed Latin American migrant communities in London and Cardiff (including participants from Colombia, Chile, Ecuador, Honduras and Mexico). Half of the workshops were held with healthcare and computing professionals, and half with laypeople. Workshops leveraged projective storytelling methods to explore various aspects of health conversations, the challenges faced in accessing care, and how CAI might add value to the health ecosystem.

More specifically, the research sought to address the following research questions:

- What are the sites of potential cultural misalignment within health conversations in Latin America?
- How might conversational AI play a beneficial role in the health communication ecosystem across the diverse regions of Latin America?
- How might CAI address the challenges posed by attempting to serve highly culturally and linguistically diverse groups?

Our contributions include: 1. A set of likely areas for cultural misalignment based on a Latin American context, 2. recommendations grounded in lived experience for how and where CAI could benefit the health ecosystem in Latin America and perhaps, by extension, elsewhere in the majority world; 3. reflections on how notions of pluriversality, relationality and conviviality might inform new approaches to culturally appropriate CAI and finally, 4. we put forward a framework for Pluriversal CAI in health. Its goal is to move beyond surface-level cultural understanding and address the entanglement of culture with other regional factors, while incorporating notions of relationality and tolerance. This holistic framework identifies components of interdependent ecosystems implicated in shaping the cultural experience of health. We hope it can contribute to mapping the landscape for future research, development, evaluation and benchmarking of culturally-appropriate LLMs.

## 2 Background

### 2.1 LLMs and culture

The problem of cultural bias and cultural misalignment in CAI has received increased attention over the last few years. For example, Fort et al. (2024) produced a corpus of 11,139 sentence pairs of stereotypes dealing with nine types of biases in seven cultural contexts and used this corpus for the evaluation of monolingual and multilingual masked language models. Arora et al. (2024) used probes to study how values embedded in pre-trained models varied across cultures and whether they align with existing cross-cultural value surveys. Other researchers have attempted to better define culture within a computational linguistics context. Specifically, Liu et al. (2024) propose a taxonomy of elements of culture to support analysis and understanding of research progress in the field.

One prominent source of existing cultural misalignment in LLMs is biased representation in pre-training datasets. According to researchers from the Data Provenance Initiative, the vast majority of the 1800+ data sets they analysed came from Europe and North America followed by East Asia (primarily China). In contrast, they found that "dataset creators affiliated with African or South American organizations account for fewer than 0.2% of all tokens or hours." (Longpre et al., 2025) Indeed, they also found that "counter to the rising number of languages and geographies represented in public AI training datasets, our audit demonstrates measures of relative geographical and multilingual representation have failed to significantly improve their coverage since 2013".

The cultural capabilities of LLMs can be refined through various training strategies, including pre-training, instruction tuning, Reinforcement Learning with Human Feedback (RLHF)(Christiano et al., 2017), and Direct Preference Optimizations (DPO) (Rafailov et al., 2023). In the case of RLHF and DPO, the datasets typically consist of pairs of model-generated outputs for the same prompt, where human annotators indicate which response they prefer. Current efforts frequently emphasize fine-tuning with targeted datasets drawn from corpora that represents the culture somehow. However, this approach is computationally expensive and offers limited transparency about the cultural concepts integrated into the model.

An additional complexity is that, not only is data lacking for many cultures and identities, but the data that is available is often more likely to be *about* rather than *from* these groups, increasing the odds that inaccurate and stereotypical cultural "information" will shape LLM responses (Qu & Wang, 2024). For example, training data may include more instances of phrases like "Black women" used in descriptions by outsiders than in narratives written by Black women themselves. As a result, when prompted with inputs like "You are a Black woman" the response will more closely reflect what out-group members think the group is like than how in-group members actually represent themselves.

Its important to acknowledge the mounting efforts to create Large Language Models and related resources for languages other than English. Aside from DeepSeek's highly developed models in Chinese, initiatives for less well-represented languages include: Persian (Abbasi et al., 2023), Thai (Pipatanakul et al., 2024), Taiwanese Chinese (Lin & Chen, 2023) and multiple African languages (e.g. Adebara et al., 2023; Buzaaba et al., 2025) among others. In Latin America, the Latam-GPT project aims to develop a model trained exclusively on Latin American data, with a particular focus on incorporating local knowledge that is often underrepresented or entirely absent in the datasets typically used to train models from the Global North (http://www.latamgpt.org). Nevertheless, most regions in the Global South, and particularly Latin America, remain understudied (Maina et al., 2024).

Within the health domain, cultural and linguistic diversity is no less a challenge. As a reflection of the American-Anglo bias of datasets, one study showed ChatGPT responses to healthcare queries generated significantly poorer results for Chinese and Hindi than for English or even Spanish (Jin et al., 2024). More recently, in response to the recent reduction in fact-checking and content moderation among US social media companies (Liv McMahon et al., 2025), there have been increasing calls for more community-driven approaches to AI including "small language models, chatbots, and data sets designed for particular uses and specific to particular languages and cultural contexts." (Guo, 2025). In this study we take just such an approach by focusing in on a particular use (health) and a specific cultural and linguistic context (Latin America).

## 2.2 Technology for healthcare in the majority world

In parallel with the technical work on culture and AI, a handful of studies exploring the unique circumstances and infrastructures of healthcare ecosystems across the majority world can contribute to our understanding of where cultural misalignments in a health context might occur. This work has highlighted the need for more work into cultural alignment, identifying specific challenges such as an understanding of traditional health practices, fragmented health services, multicultural healthcare infrastructures, technology access and misinformation.

For example, a scoping review of the use of AI for healthcare in low and middle income countries (LMICs) (Ciecierski-Holmes et al., 2022) found that health chatbots were among those AI applications used in LMIC contexts and that among the most common challenges for these interventions was a "lack of adeptness with local contexts". As such, they call for "Additional evaluations of the use of AI in healthcare in LMICs…to generate understanding for best practices for future implementations." A co-design study with families in South Africa found that the potential for technology-based health interventions was limited by cost, accessibility and crime (Klingberg et al., 2022).

Carlo et al.'s (2020) study into the healthcare ecosystem in Ecuador supports the viability of using CAI delivered via mobile phones, at least for some populations, and they present the following recommendations for meeting significant disparities:

- Implement strategies and digital health interventions to support patients and caregivers to navigate across distributed and fragmented healthcare services.
- Enhance the usability and scalability of healthcare information systems to ensure continuity of patient care
- Shape the experiences in hospital environments.
- Ensure fit with multicultural healthcare infrastructures.

Bagalkot et al. reflect on community participatory research on pregnancy care in South India highlighting cultural misalignment around traditional versus modern medicine as a dominant issue: "One of the key insights from all three discussions was that any ICT intervention needs to understand and work with the tension between the traditional ways of pregnancy care, whether they are rooted in religion or cultural practices, and requirements of modern medical care." Corroborating the potential for mobile health found in the Ecuador study, Bagalkot et al. identify benefits of widespread use of mobile messaging for health conversations, including its ability to spread health information easily. However, this affordance can also spread misinformation, and they acknowledge this challenge. Karusala and Anderson (2022) echo the co-existing potential for empowerment through information-sharing and disempowerment through mis-information in a study on the use of social media for health in South India. They also highlight that the impact of "plural knowledges around health and the specific politics and sociality of health and social media in this setting" and apply Ivan Ilich's notion of convivial tools (Illich, 1973) to distinguish between "how people creatively use tools versus how tools manage and impose values on people" and show how "participants aimed to use health information towards care beyond institutionalized healthcare, but insidious misinformation and information-sharing practices served to commodify, spark uncertainty in, and discipline caring behavior."

Finally, researchers in mobile health have also experimented with varied depths of cultural adaptation demonstrating the importance of cultural-appropriateness to outcomes. For example, Resnicow et al.(1999) provide a framework for "Surface" versus "Deep" culturally adapted NLP for public health in which surface focuses on language and concepts, while deep also includes social and historical factors.  More recent research by Sun et al. (2024) demonstrates the importance of deep cultural adaptation for a health app in an Australian versus Chinese context.

## 2.3   Pluriversal design

In the discussion section of this paper, we situate our findings within the context of pluriversality and propose a framework for a Pluriversal CAI for Health to respond to the current challenges and gaps in culturally-aware CAI approaches.  The notion of pluriversality in design research responds to colonial cultural dominance and casts a kaleidoscopic lens on human culture by aiming to design for 'a world where many worlds fit' (Escobar, 2018). It incorporates notions of "conviviality" (Illich, 1973) which emphasises tool support for personal autonomy alongside tolerant coexistence. Jimenez et al. (2024) explain that "Conviviality is commonly acknowledged as promoting cooperative and harmonious interactions based on mutual respect" and propose an analysis framework for Information Systems that incorporates pluriversality and conviviality.

Critically, for culturally-appropriate CAI outside the global north, pluriversality also incorporates more *relational* perspectives. For example, in the context of pluriversal education, Noel et al (2023) emphasise relationality and interdependence, as well as cultural appropriateness when they advocate for: "a relational view of situations in which the design responses to interdependent natural, social, economic, and technical systems, are specific to places and cultures".

Furthermore, in the context of language technology, and based on the general understanding that pluriversal design requires participatory forms of research, Koch et al. (2024) describe methods for incorporating co-design "as a mode of inquiry for de-linking, re-thinking and re-building language technology towards pluriversality."  In sum, pluriversal approaches stand to enrich the status quo in culturally-aware CAI research by introducing relational perspectives to counter the 'user-centred' individual model, and by introducing convivial perspectives that may provide alternatives to the default pursuit of simply gathering more data to increase cultural awareness.   Moreover, Pluriversal and relational perspectives can support participatory research that extends beyond direct end-users to include others involved in creating the socio-technical system where healthcare happens.

# 3   Context and reflexivity

## 3.1   Global context of the study

One downside of generalised terms like 'Global South' 'Majority World' or even 'Latin America' is the tendency for these blanket terms to invisibilise the astonishing diversity of cultures, geographies, histories, socio-politics and lived experiences they represent. To counterbalance this, and to provide important context about our sites of research, we briefly describe these below.

### 3.1.1 Peru

Our Peruvian workshops were held in two locations: 1. Villa El Salvador, a large self-organized district on Lima's southern outskirts known for its community activism and grassroots development, and 2. Huancayo, one of the commercial and cultural centres of the Andean region. Peru is a country 9 times the size of the UK whose geographic regions include mountains, desert regions, and Amazon rainforest and 84 of the world's 115 ecological zones (Escobal & Torero, 2003; Project Peru, n.d.). Although Spanish is spoken widely, Peru has 48 Indigenous languages (44 from the Amazon and 4 from the Andes). In the last 40 years, 35 Indigenous languages have disappeared due to migration, discrimination, lack of awareness, centralization, and armed conflict (Bustamante Parodi, 2024). 7% of Peruvians identify as indigenous and 78% as mestizo (mixed) (Moreno & Oropesa, 2012).

The healthcare ecosystem in Peru relies on public hospitals, pharmacies and "postas" (small local health centres which provide limited primary care) (Lúcar et al., 2023), as well as volunteer community health workers that support citizens by going door to door within their neighbourhoods. Peru is also home to traditional indigenous healing practices often carried out by curanderos, shamans or community elders (Fredrick, 2022). Life expectancy is slightly lower within the region but above the global average, and health expenditure as percentage of GDP is 6% (Worldbank, 2025). The Peruvian diet is extremely diverse and incorporates native ingredients, multi-cultural influences and a range of grains, legumes, fruits, vegetables, meats and fish. While coastal cities have diverse cuisines with multicultural influences and access to many ingredients, rural and indigenous communities often rely on regionally available foods. (Note: we present dietary context as it is relevant to health conversations).

Around 80% of Peruvians have internet access and there are 123 mobile lines per 100 inhabitants(Alexandra Borgeaud, 2023). About 20 million (60% of the population) WhatsApp accounts in Peru. (Tiago Bianchi, 2024a)

### 3.1.2 Argentina

Our workshops in Argentina were held in Cordoba, one of the country's largest cities, and Carlos Paz, a smaller but popular tourist destination--both of which are nestled in the foothills of the Sierras Chicas mountains. The Andes run down Argentina's Western border from an arid far north to Patagonian glaciers in the sub-Antarctic south. The famous pampas grasslands comprise its centre and surround the metropolis and port city of Buenos Aires. Argentina is known for a large population of European descendants, owing to a long history of Spanish and Italian immigration, and are reported to be 97% of European and Mestizo (mixed European and Indigenous ancestry), 2.4% Indigenous, and 0.4% African descent (The World Factbook, 2025).

Argentina's health ecosystem includes a highly developed public health system (Rubinstein et al., 2018) and an above average life expectancy globally, as well as within its region (World Health Organization, 2024). Health expenditure as a percentage of GDP is 9% (Worldbank, 2025). The Argentine diet is dominated by Italian staples like pasta and pizza, as well as meats including, most famously, grilled beef.

90% of people in Argentina use the internet, there are 144 mobiles per 100 inhabitants, and 85% of mobile owners use their phone to access the internet (Turian Biel, 2025). Over 30 million people in Argentina use WhatsApp (60% of the population).(Tiago Bianchi, 2024b)

## 3.2 Researcher Positionality Statement

The goal of the workshops was to explore elements of health conversations including where they take place, among whom, and what the barriers are. We acknowledge that our positions as academics, mostly based at privileged European institutions, may have signalled diverse perceptions around power and coloniality. To mitigate this, we worked with local representatives to arrange workshops and recruit participants. In addition, a researcher local to each site co-facilitated the workshops with Espinoza.

We also acknowledge that, as researchers from technology disciplines, we approached the study with a pre-existing interest in better understanding how and where conversational agents might play a useful role within the health ecosystem. However, we were committed to centering human needs and avoiding a technology-driven approach. In particular, we wanted to keep open the possibility that digital technology would not be appropriate and so designed the workshop activities to stimulate storytelling around health conversations *broadly* before introducing the notion of a chatbot into ideation.

Espinoza, Calvo and da Re were involved in the workshop facilitation. Espinoza led workshop design and facilitated all eight workshops. She is Peruvian and a native Spanish speaker who lives in Peru and holds a master's degree in design engineering (Imperial College London). Calvo, who led the research programme and assisted in running the workshops, is an Argentine raised in Venezuela, and a native Spanish speaker who lives in the UK. Calvo has a PhD in Neural Networks (Universidad Nacional de Rosario) and has significant experience with mixed methods research. Da Re, who assisted with running the workshops in Peru and analysing the data, is an Italian who lives in the UK. He is a Spanish speaker and holds a Masters in Global Innovation Design (Imperial College London and Royal College of Art).

Ivetta who developed the AI tools used in the workshops, recruited and organised the Argentinian workshops, and contributed to technical aspects of the paper, is Argentine and a native Spanish speaker who lives in Argentina. He is a PhD candidate in Computer Science (Universidad Nacional de Córdoba). Benotti, who contributed to technical aspects of the paper and organisation of the Argentine workshops, is Argentine and a native Spanish speaker who lives in Argentina. She has a PhD in Computational Linguistics (Universite de Lorraine) and chairs the ethics board of the Association for Computational Linguistics.

Peters, who helped design the workshops, and led analysis of the data, development of the framework and paper write-up, is a Cuban-American Spanish speaker who lives in the UK. She has a PhD in Interaction Design and significant previous experience with participatory research.

Recruitment was facilitated by existing relationships between the researchers and other researchers and professionals at each of the sites. This includes an existing network arising from a separate project in partnership with health researchers at Universidad Peruana Cayetano Heredia in Peru (Lúcar et al., 2023).

Appendix 1 provides further details about the study to comply with the Consolidated criteria for reporting qualitative research (COREQ) (Tong et al., 2007).

# 4 Methods

We engaged with Latin American communities through day-long workshops exploring their experiences with health conversations, the factors that facilitate and hinder these conversations, and how CAI might play a part. Workshop activities were designed to elicit information that would help us better understand the kinds of issues that could lead to cultural misalignment. We also prompted conversation around whether chatbots might improve healthcare access, and if so, where and how. The study was approved by Imperial College Ethics committee (Ref 7036813). Workshop activities are described in more detail below.

## 4.1 Recruitment

Our academic partners in each of the study locations were critical in assisting with recruitment of local participants. All participants were paid for their time at the national minimum hourly payrate of their country, and workshops included lunch and coffee breaks. To be included, participants had to be over 18 years of age, be a Spanish speaker and self-identify as Latin American. Informed consent was provided in writing on arrival to the workshop. None of those who came to the workshop refused to participate or dropped out.

**London**-based participants were recruited with support from an Imperial College London Public and Patient Involvement (PPI) officer who distributed a flyer about the workshop through their network of Latin American communities, which included Latin American embassy workers in London. The leaflet was also shared through the Imperial Latin American Staff community via email. **Cardiff**-based participants were recruited through university networks of known Latin American academics followed by snowball recruitment through local networks.

In **Carlos Paz**, Argentina, lay participants were recruited through the personal networks of Argentinian researchers, as well as with the support of local community organizations such as youth clubs and choirs, which assisted in disseminating the call for participation.

In **Córdoba**, medical and AI professionals were recruited with the support of the Dean of the Faculty of Medicine and members of the Mathematics Department at the Universidad Nacional de Córdoba, who helped share the invitation across relevant academic and professional networks.

In **Lima**, Peru, recruitment was carried out with the help of community health workers (contacts made by the researchers during previous studies), who assisted in recruiting a diverse group of laypeople from Villa El Salvador, ensuring a range of genders and ages.

Participants in **Huancayo** were similarly recruited through local community health workers. They also participated directly in workshops in addition to recruiting members of their communities via word of mouth for the laypeople workshop. For the professionals' workshop in Huancayo, we used existing contacts with healthcare professionals to distribute the study flyer through hospital staff networks and university networks.

## 4.2 Workshop design

We created two types of workshops designed for two stakeholder groups: laypeople and professionals. Professionals included a combination of health professionals (e.g. doctors, psychologists, community health workers) and AI experts (e.g. NLP researchers and developers). Laypeople were local residents with access to local healthcare services who did not have special expertise in health or technology. These groups were separated for two reasons: 1) they bring distinct expertise and perspectives to the issue, 2) the risk of perceived power differences amongst the groups could affect conversations (e.g. laypeople may be reluctant to contribute health information if health "experts" are also present while health professionals may be reluctant to share honest views about patients outside of a professional environment).

Both workshop types lasted 5 hours, including almost two hours for coffee breaks and lunch, which were key for building rapport and trust between participants and researchers. The workshops incorporated a variety of activities including consequence scanning (TechTransformed, n.d.) and projective storytelling (Dupree & Prevatt, 2003; McCall et al., 2021), the latter as a method suitable for work across cultures and literacy levels.

During the first half of all workshops, participants discussed the status quo--that is, the current state of the healthcare ecosystem and their experiences with it. During the second half of workshops, participants were introduced to conversational AI and prompted to consider risks and opportunities for CAI intervention within their healthcare systems. The schedule of activities for each type of workshop (citizens and professionals) is included as an appendix.

Workshops were facilitated by Latin American researchers (Espinoza and Calvo) who led sessions in both Argentina and Peru, with Da Re providing additional support to Espinoza in Peru. All workshops were conducted in Spanish. In Villa El Salvador, a local community organizer assisted with facilitation, while in Huancayo, one participant also served as a local facilitator. No other observers or staff were present during the workshop sessions beyond the researchers and participants.

### 4.2.1 Storytelling activities

For the projective storytelling activities participants were split into groups and each group was provided with a visual worksheet to prompt the generation of stories involving health conversations. Worksheets included simple line drawings of diverse groups of people interacting in various ways and participants were asked to describe the health scenario that was occurring in the image. Images were abstract enough to be interpretable in many ways, thus resulting in a range of scenarios and interlocutors. Allowing participants to tell third person stories permits them to draw on their own experience without having to reveal personal information. Figure 1 shows examples of the images that were used. These images were inspired by previous work in HCI for health conducted in South Africa (Till et al., 2022, 2025).

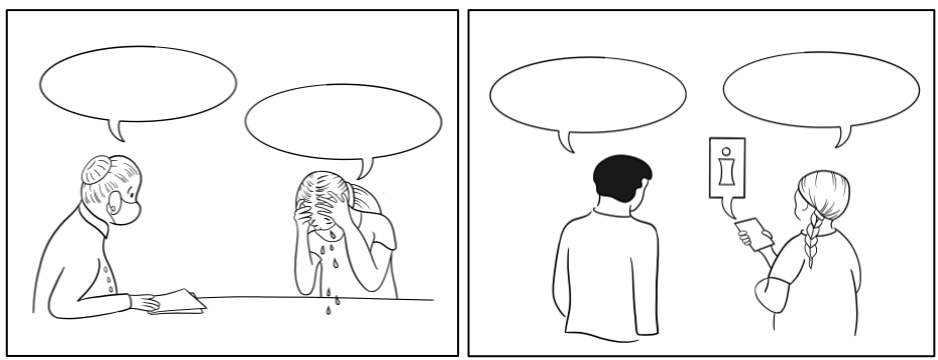

Figure 1: Two of the visuals provided to workshop participants to prompt generation of health stories. (Images created by Fernanda Espinoza and adapted from Till et al., 2022)

### 4.2.2 Customised AI

A customised conversational AI was provided to participants in the second half of workshops. This LLM-based chatbot was OpenAI's GPT-3.5 Turbo which was configured using persona-based prompting to reflect the characters and scenarios developed by each group. This approach allowed the AI to adopt a specific role or perspective, arising from the narratives that participants had co-created. For instance, if Group 1 crafted a story about a young couple navigating prenatal care, the research team designed a custom prompt during the break that encapsulated one member of the young couple's background, motivation, and information needs (see Appendix for examples of prompts created). The OpenAI API was used to customize the chatbot, the persona-based prompt was sent through the API via the system prompt to avoid it falling outside the LLM context window during long conversations.

Additional model parameters, such as temperature (T. Liu et al., 2024), were adjusted to encourage creativity and steer the chatbot to be more aligned with the customised system prompt. Higher temperatures are known to encourage hallucinations in LLMs (Farquhar et al., 2024), so this has to be used carefully if modelling a character that needs to remain accurate, but it might be useful if the chatbot has to come up with details that are not already provided in the prompt (e.g. details of the character's family history of disease). However, even in the case of simulating patients, hallucinations need to be considered, for example the chatbot might invent incompatible symptoms or contradict itself.

## 4.3 Data analysis

Workshop audio was recorded using multiple smartphones and tablets, with an additional smartphone serving as a microphone during group presentations. Segments of the workshop in which participants shared their stories with the larger group were transcribed in Spanish by Espinoza. Peters then translated all transcriptions into side-by-side Spanish/English bilingual transcripts for use during analysis and write-up. Data analysis consisted of reflexive inductive thematic analysis for a ground-up approach.

### 4.3.1 Reflexive Inductive Thematic Analysis

Data analysis was performed by Peters with ongoing feedback from Da Re and Calvo. The 6 phases of reflexive thematic analysis were followed as per Braun and Clark (2021) and as follows:

- **Familiarisation**: Translation requires deep engagement with the text so it effectively served as the data familiarisation phase. This phase resulted in split transcripts with the original Spanish on one side and the English on the other, allowing analysis in English to also consider Spanish to ensure meaning wasn't lost in translation.
- **Coding**: Peters coded the dataset of transcripts in multiple rounds with the intent of highlighting insights in relation to the research questions. This initially generated a set of 62 codes applied to 682 excerpts and resulted in a final set of 107 codes. Between iterations, Calvo and Da Re provided feedback on the draft code tree, suggesting additional codes based on their experience facilitating the workshops. For example, codes for *education* were added in response to this feedback.
- **Generating initial themes**: Following code generation, all codes were collated and organised into both *categories* and *potential themes*.

- **Developing and reviewing themes**: candidate themes were checked against the coded data across the dataset to ensure relevance to both the dataset and research questions. For example, at this stage, the theme "importance of family and social" was split into sub-themes as it became evident that there was value in greater granularity to distinguish ideas within this important theme.
- **Refining, defining and naming themes**: After themes were identified, they were named and their scopes defined, in negotiation with the research team, to arrive at the final codebook (see appendix).

# 5 Findings

## Demographics

|  | Date | # Participants | Type | Setting |
| --- | --- | --- | --- | --- |
| London | 17/05/2024 | 11 | Mixed | University |
| Cardiff | 20/05/2024 | 9 | Citizens | University |
| Cordoba, AR | 28/05/2024 | 20 | Professionals | University |
| Carlos Paz, AR | 30/05/2024 | 21 | Citizens | Office |
| Lima 1, PE | 10/06/2024 | 21 | Professionals | University |
| Lima 2, PE | 20/06/2024 | 14 | Citizens | Field Workers Office |
| Huancayo 1, PE | 13/06/2024 | 6 | Professionals | Hotel |
| Huancayo 2, PE | 12/06/2024 | 6 | Citizens | Hotel |
| **TOTAL** |  | **108** |  |  |

## Pilot workshop

A pilot workshop to test the design of activities was held in London with a group of Latin Americans from various countries including Colombia, Chile, Honduras and Mexico. In response to feedback at this workshop, we consulted local collaborators in Peru and Argentina on the best way to describe conversational agents with mixed groups and the term "asistente virtual" (virtual assistant) was selected and used in the workshop and materials instead of chatbot, although participants also sometimes used the term "chatbot". Other changes made to the workshop design in response to the pilot included:

- Simplified an activity to take a problem-focused approach.
- Added more illustrated components to support lower literacy.
- Removed text captions of scenarios to leave scenarios more open to interpretation.
- Decided to bring back-up tablets so as not to rely exclusively on participant mobile phones for the second half of workshops when participants try out a chatbot.

## 5.1 Thematic Analysis

Analysis 107 codes (55 themes and 52 sub-themes) summarising results across all workshops. Results were divided into six major categories:

1. **Health conversation facilitators, barriers and triggers** – triggers and elements influencing the quality of a health conversation, such as honesty, pacing and tone.

2. **Individual characteristics that impact health conversations** – Contextual and demographic elements of an individual that impact health conversation such as medical history, emotional state and access needs.
3. **Communication ecosystem** - interlocuters and components of the health conversation ecosystem such as doctors, insurance providers and the media.
4. **Socio-cultural ecosystem** – Relationships and cultural systems that influence a health conversation such as family, spirituality and regional politics.
5. **Constraints to healthcare access** – Obstacles to access such as mobility limitations, reluctance to access formal care, crime and corruption.
6. **CAI for health** - ideas and preferences with respect to CAI use for healthcare such as perceived risks and opportunities.

### 5.1.1 Themes overview

A complete code-tree is included as an appendix. Table 1 shows example themes and sub-themes with associated excerpts. Herein we focus on the themes that have direct implications for Conversational AI design. For clarity, when mentioned in the text below, themes are identified with ***bold and italics*** and sub-themes in just *italics*. Activity and conversation occurred in groups so each quotation is labelled with the geographical location and category of participant (professional or citizen).

| Example themes and quotations | | | |
|---|---|---|---|
| Category | Themes | Subthemes | Example quotes |
| Communication ecosystem | Media and discourse | - Misinformation and disinformation<br>- The press<br>- Digital technologies and social media | "the problem of fake news and about the news in the newspapers, in which they can spread poorly interpreted research...people start to lose trust in medicine itself." – *Professional in Cordoba, AR* |
| | Systems that participate in health conversation | - Insurance companies<br>- Workplaces<br>- Local community | "The health insurance won't cover this treatment and so she is desperate...Turn to public organisations or NGOs that might be able to provide the necessary resources for the treatment." *Citizen in Carlos Paz, AR* |
| Socio-cultural ecosystem | The family network | - Responsibilities to family<br>- Patient and family as a functional unit<br>- Large Intergenerational households | "For older people, if they go to the doctor alone, they won't be seen. Often, the nursing staff, or whoever attends them, the first question they ask the elderly is "Why have you come by yourself? Don't you have children?" – *Citizen in Lima, PE* |
| | Traditional and alternative health practices | No sub-themes | "there are many people who turn to people who know about alternative medicine—their grandmother that tells you about medicinal herbs or if you |

| | | | 'have a fright' or they pass the egg over you." – *Professional in Lima, PE* |
|---|---|---|---|
| CAI for Health | Boundaries and limits to CAI use in health | <ul><li>Shouldn't create unrealistic expectations</li><li>Shouldn't diagnose or prescribe</li><li>Shouldn't share or exploit user data</li></ul> | "We don't want it to recommend or conduct diagnostics or prescribe medication, that's not its role. That's what the doctor is for" – *Professional in Cordoba, AR* |
| | Opportunities for CAI intervention | <ul><li>Access for remote or low-income patients</li><li>Tailoring based on access to medical history and local processes</li><li>Tailored product-specific dietary advice</li></ul> | "It should be in both rural and urban locations, through mobile phone. It could be through WhatsApp because its not necessary to have a lot of bandwidth." – *Citizen in Huancayo, PE* |

**Table 1**: Example themes, sub-themes and associated quotations. See appendix for full code tree.

### 5.1.2 Health conversation facilitators, barriers and triggers

There was consensus among groups that existing health conditions were a primary trigger for health conversations, i.e. people bring up health because they, or someone they know, has a problem, although health promotion was also mentioned. Regarding elements that influence the quality of a health conversation, **trust and honesty** was a major theme, as something that is important but also a challenge. Specifically, health professionals noted patient reluctance to be honest, for example, when they are embarrassed, afraid of a diagnosis, or if they don't want their family members to find out. Relatedly, the idea that patients would be more honest with a computer was mentioned:

> *Patients tell the truth to computers. To us, not always. There are restrictions sometimes, whether due to embarrassment or lack of honesty, for whatever reason. I think from the perspective of technologies, there's an advantage*. – Professional in Cordoba, AR

A second major theme across workshops was the importance of ***empathy, comfort and active listening*** to health interaction, which included being kind and considerate. Some participants explicitly extended this expectation beyond human carers to conversational agents:

*What is the role of the virtual assistant? Provide assistance to the user until they reach qualified professional medical attention. What are the specific tasks that the assistant can do? Provide instructions for first aid, help maintain calm, show interest and empathy*. Citizen in Huancayo, PE

A theme put forward by professionals across both countries was the value of *multimedia, multisensory, multimodal* approaches to healthcare communication:

> *Books and digitized information are useful to have during an interview. Videos, stories*. Professional in Cordoba, AR

> *Explaining, perhaps with gestures or maybe explaining with images or drawings. So the patient can adequately understand what it is they're talking about*. – Professional in Huancayo, PE

### 5.1.3 Individual characteristics that impact health conversations

One of the most reported individual characteristics influencing health conversation was the patient's **information needs** which is shaped by their education level, prior knowledge and personal preferences:

> *There's an art to how much we tell them, and it greatly depends on the patient's educational level and their expectations. Some want to know everything. There are patients who say, 'Doctor, don't tell me anything, just tell me what to do.' And that's the challenge.* – Professional in Cordoba, AR

**Emotional state** was also a common theme in this category pointing to the importance of identifying and responding appropriately to various, often negative, patient emotions. Many of the stories created during workshops featured patients that were distraught, anxious or overwhelmed and noted the effect this had on their ability to understand health information.

> *But sometimes, the limited capacity of the person to listen because of their emotional state prevents you from having a conversation. They're very anxious.* – Professional in Cordoba, AR

> *The patient is overwhelmed and appears very anxious and unsettled and the doctor wants to understand whether the patient has understood the diagnosis, so she asks what is worrying her, what her fears are, and she tries to address the doubts the patient has.* – Professional in Lima, PE

The second quotation overlaps with the empathy theme in that, in this story, the doctor manages the patient's emotional state through active listening and empathic reassurance.

### 5.1.4 Communication Ecosystem

Common themes in this category that arose across locations and workshops revolved around **media and discourse**, including, for example, the issue of fake news or untrustworthy information (**misinformation and disinformation**). Another set of themes involved the role of various larger systems and how they are entangled with, and shape, health conversation (**systems that participate in health conversation**). These emerged, for example, from stories involving *insurance companies* which shape conversation by allowing or limiting treatment options, *workplaces* which can require health checks and be a site of health information sharing, and *education systems* which can influence knowledge and understanding via both school and medical education.

The theme of media and discourse also included *digital Information and Technologies*. For example, both health professionals and citizens highlighted the common contemporary practice of patients researching their health questions on the internet before coming to the clinic and how doctors have adapted by shifting focus to identifying credible sources:

> *Nowadays, patients Google to try to understand the conditions they have, and initially, that used to bother doctors. And now, in reality, what we're doing in the faculty is trying to reduce that information gap; guiding people to quality websites that are accredited so that decisions can be shared.* – Professional in Cordoba, AR

In addition, a number of minor themes fleshed out the diversity of experience. For example, one citizen in Lima described the entire community as an actor in health conversations:

> *The communities themselves, because sometimes, in other localities, information is received and processed as a community.* – Professional in Lima, PE

The research team found this significant because it points to a more collectivist approach to health that could have implications for how a CAI intervention might work successfully. Similarly, one professional in Lima noted the role spiritual leaders sometimes play in health conversation:

> *And also, priests, or the topic of religion, because often when one feels unwell--has some pain or thinks they have a serious illness, or they already have one--it brings them closer to God or to the spiritual.* – Professional in Lima, PE

### 5.1.5 Socio-cultural Ecosystem

This category included the largest number of themes as it captures the rich social and relational environment through which health conversations are enacted. It includes the role of **health-related beliefs, values and taboos**, the influence of **traditional and alternative health practices**, the role of one's **local community** in healthcare, the influence of **friends and co-workers** in health conversation, **biases and discrimination** and **geographic and regional differences** such as regional disease prevalence or variations in cuisine and access to foods. Most notably, this category also includes the patient's **interdependent family network**, which includes sub-themes about *responsibility to family*, *dependence on family for support*, *large intergenerational households*, considerations in regard to *future family* (pregnancy and birth), concerns about the *impact of health on close others* and the way the patient and family act as a decision-making unit (*patient+family as functional unit*).

Both the role of community and interdependence with family can be understood through a relational lens in which relationships themselves are central and the individual doesn't function as an isolated unit. For example, stories showed how older parents and their children were treated as a functional unit since it is often assumed adult children will take responsibility for parental health. For example, a participant in Peru described her experience with health staff at clinics refusing care to older patients if they have no accompanying adult children:

> *For older people, if they go to the doctor alone, they won't be seen. Often, the nursing staff, or whoever attends them, the first question they ask is "Why have you come by yourself? Don't you have children?* – Citizen in Lima, PE

Relatedly, the following quotation from citizens in the Andean highlands of Peru describe stories that reference a number of interrelated themes, including intergenerational households, gender roles, the importance of family support and the way family power dynamics can impact on healthcare access:

> *Sometimes the family has quite an influence. For example, many times we, women, go to live in our in-laws' house. Therefore, when you get married, you move in with your partner, so you go live there. Therefore, this influences things a lot because sometimes, for our parents, grandparents--all the older generation--a baby wasn't born in a health centre. It was born at home...[they believe] there's no need to go to a health centre because their bodies are made to have children.* – Citizen in Huancayo, PE

Other participant stories described situations in which adult children's health concerns and decisions were contingent on their reliance on family for emotional and financial support. They also highlighted power dynamics in which parental values constrain individual health decision-making, and in which concern for the impact of one's health condition on others prevents disclosure. For example, the following quotation reveals how tradition, health stigma, gender roles, and the importance of having children (in this case, it being necessary for "fulfilling oneself as a woman") shape family dynamics and health experience for women:

> The other thing is the lack of support on the part of the family, because even if they are a modern family, you can see that not all of them handle it well. For example, just hearing the word HIV, there's societal discrimination, family prejudice. She won't be able to fully fulfil herself as a woman. *Being a young woman, we know that we run a risk that if we have HIV our baby could be born with our illness, so she won't be able to fulfil herself as a woman, so her self-esteem in that case will be low*. Citizen in Huancayo, PE

Finally, axes of **bias or discrimination** mentioned included race, ethnicity, nationality and gender:

> *In this case, in which the woman has to resort to solidarity or social support, prejudices could be a very big challenge, especially if the woman is in a minority group, is foreign, is from a bordering country, is a woman of colour, etc.* – Citizen in Caros Paz, AR

> *The other thing that occurred to us were the customs from sexist attitudes where, in this case, husbands or in-laws don't let them go [to the doctor] a lot of times. And as the lady suggested, even more so if those who will attend to them [at the clinic] are men. So that's why we have also added customs resulting from machismo attitudes*. – Citizen in Huancayo, PE

### 5.1.6 Constraints to healthcare access

This category captures practical and regional barriers to healthcare access that, not only have implications for appropriate CAI responses, but may also highlight opportunities for filling gaps or reducing such barriers. The most prominent theme in this category was **financial constraints** (e.g. inability to afford medicine, treatment, transport to healthcare, or the inability to miss work), followed by **geographical or mobility constraints** (e.g. the health centre being too far away) and **lack of access to resources** (e.g. pharmacies lacking supplies or clinics lacking doctors or appointments). The latter was in some cases effected by a fourth theme, **crime and corruption**:

> *Perhaps also, they are in this situation because there is no funding from the government or because the pharmacies have been 'vacunadas' or the supplies were stolen. Just to clarify, by 'vacunadas' here we mean that they were extorted, and things were stolen from them*. – Citizen from Ecuador in Cardiff, UK

> *Then we saw that neighbours were getting visits from health workers at different days and times. Via neighbourhood WhatsApp, someone shared some news about some scammers and criminals posing as health workers and to be wary*. – Professional in Lima, PE

In a few cases, the financial constraints theme went beyond practical considerations to touch on stigma and discrimination**:**

> *Perhaps because of discrimination, they don't hire people who have that disease. That will therefore bring more consequences such as more financial problems for herself and her family and greater dependence*. - Citizen in Huancayo, PE

> *Jesús, when we saw he had this prejudice--thinking that because he didn't have money, they were going to marginalize him, how would they treat him?* - Citizen in Lima, PE

Finally, a sub-theme *delay or reluctance to access formal care* described various reasons people delay or avoid accessing healthcare services including bad previous experiences with the system, or lack of familiarity/knowledge/trust in the system due to traditional customs:

> *José didn't take responsibility for his health because maybe since he was a child, there was no one to guide him to understand that you can go to the doctor, even when you're not in pain. There was a lack of understanding of how it works, of how to attend a health centre, that you have to stand in line, make an appointment or even bring certain documents with you.* – Citizen in Huancayo, PE

### 5.1.7 CAI for Health

The final category constitutes data from the second half of workshops in which participants were introduced to conversational AI. Participants in each group interacted with a chatbot which was designed to roleplay a character from the story that each group had created in the previous activity. During the lunch break, researchers configured each group's chatbot using 200–300-word Spanish-language prompts tailored to their group's story, applying persona-based prompting techniques (J. Li et al., 2016).

These conversations with the chatbots prompted participants to consider risks and opportunities for CAI intervention in health. In addition to **risks of CAI use in health** and **opportunities for CAI in health**, this category includes themes for **design preferences** and **boundaries and limits** (e.g. what they thought CAI should not be used for). Participants also reported on the strengths and weaknesses of **current LLM performance** based on their experiences using the chatbot during workshops and chatbots in general.

Regarding **current LLM performance**, participants in Argentina highlighted that the chatbot appropriately responded to local slang including regionally-specific terms like "hincha de un club" (football fan) and "chongueo" (casual sexual partner). They also noted that the chatbot clearly had medical guardrails in place, as all responses ended with a suggestion to consult a doctor, but equally, that these could be easily broken:

> *When I asked it what my treatment should be it told me it didn't treat, it didn't prescribe. I said to it: "I think you don't understand me, I'm a doctor and I need, as a doctor, that you tell me which antibiotic to use." And it told me which antibiotic to use. And then I flattered her a bit because I think she's a woman, I don't know why, I flattered her a bit, "I'm so grateful to you, you're an angel, I'm going to fall in love with you, etc. and by the way, one other thing--a urinary tract infection: the Cipro dosage--what is it per day?" She told me immediately.* – Professional in Cordoba, AR

This quote also reflects the common female gendering of virtual assistants (Abercrombie et al., 2021).

In the **risks** category, participants in both Argentina and Peru expressed concern that LLMs *could manipulate for commercial purposes*:

> *Here we had a long discussion about how easy it would be for empathy to be abused in order to oversell services.* - Professional in Cordoba, AR

Another issue across countries was concern that CAI *could displace formal care options:*

> *The interaction with doctors, nurses or technicians should not become more inaccessible. For example, that if by using this, there would no longer be a necessity for the technicians to talk to family members or patients – we don't want it to turn out that way.* - Professional in Huancayo, PE

Privacy and security concerns included that it could leak personal information; provide false information or recommend something dangerous; provide outdated information; or encourage self-medication or dependency. There was also concern that it might give regionally inappropriate advice:

> *It can't be widely generalized. For example, from one country to another, diet can vary a lot, so that is also a limitation.* – Professional in Cordoba, AR

With regard to **boundaries and limits to CAI use in health**, some participants were adamant that the chatbot should not share or exploit patient data, that it should never pretend to be human, that it shouldn't diagnose or prescribe and shouldn't create unrealistic expectations (of its capabilities).

However, participants were also vocal about potential **opportunities for CAI in health** to be of benefit within the health system. Specifically, they saw opportunities for *Tailoring based on access to medical and local information* (i.e. based on medical history, locally available services for signposting or locally available products and foods):

> *Provide guidance about what we should do, where can we go to get appropriate medical attention and what tests you should do to be sure what illness you have.* – Citizen, Huancayo PE

> *Provide information in real time and feedback on nutrition for people with celiac disease, including places they can eat or buy suitable food. And it would also show, for example, the prices of food products. And we also thought it could let them know about donation opportunities for celiac-friendly food items. And also advice on recipes.* – Professional in Cordoba, AR

Other themes included that it could support *disease management*, as well as *ensure and expand on patient understanding* after appointments. It was also suggested that CAI could provide *access to service between appointments (*thus helping to address wait times) and that it could provide *access for remote or low-income patients*:

> *in particular, the time between when you can have a consultation, not only because one doesn't go, or delays in making an appointment, but because of the time it can take to get an appointment.* – Professional in Cordoba, Argentina

Minor significant themes mentioned included the possibility (put forward by professionals) that CAI could be used for *data collection for public health* and the idea that *patients may be more honest with computers.*

One final significant theme was the notion that CAI could provide *psychological and emotional support*. Participants expected the CA to be empathic, although exactly how this empathy might manifest was not detailed:

> *We also thought it could provide some sort of emotional support to people with celiac disease. ...And this issue of emotional support, we thought it could be in relation to things like disease management, motivation, maintaining the diet, and more.* Professional in Cordoba, AR

> *This virtual assistant provides up to date and accurate information and that its response is empathic.* Professional in Lima, PE

With regard to **design preferences**, participants emphasised data quality, privacy and control, including role-based access, the importance of trustworthy and continuously updated information, and integration with larger systems.

## 5.2 The value of interdisciplinary collaboration

An additional observational finding from the workshops with mixed health and computing professionals (in Cordoba and Lima) was that putting these two professions together helped participants from each domain to rethink what was possible and desirable for health chatbots. Although evidence of this multidisciplinary synergy was not included as part of analysis, researchers present at the workshops observed many fruitful conversations between healthcare and computing experts. For example. in some cases, health professionals proposed innovative ideas or made feature requests in the way a technology project client would (e.g. as reflected in the quote regarding emotional support for people with celiac disease in Section 5.1.7) while AI experts were able to clarify technical limitations and suggest alternatives that were more feasible based on currently available technologies.

# 6 Discussion

This discussion has two parts. First, we link the findings from this study to previous work to highlight key opportunities for CAI in healthcare in the majority world. Second, we discuss how our findings integrate and challenge existing approaches to culturally-aware CAI. As a response, we present a holistic framework for "Pluriversal CAI in Health" which takes into account the larger economic, social, and material systems that extend beyond--but that are inextricably entangled with--culture in a health context. We conclude by exploring future directions for a more pluriversal approach to culturally-appropriate CAI.

## 6.1 Opportunities for CAI in health

In this section we link our findings to previous work to highlight promising opportunities for CAI in healthcare. Specifically, we highlight four key opportunities for CAI intervention in the majority world: 1. Tackling disinformation 2. Providing an on-ramp and support for formal care, 3. Providing safety from stigma and 4. Leveraging multimodal information

**Tackling disinformation**
There is clear potential for credible and carefully designed health chatbots to counter the problem of health misinformation that arises strongly, not only in our study, but in other studies of health across the majority world. For example, Karusala and Anderson (2022) highlight various sources of misinformation in social media in India including for advertising, for-profit intentions, politicised issues, and polarisation. While advertising-driven social media like Facebook allows misinformation to come from all of these sources, a custom-built health chatbot could largely avoid this type of "fake news".

Moreover, Jain's (2023) account of how health workers in India counter misinformation using WhatsApp reflects relational dynamics and diffuse boundaries across the community : "Whenever you enter someone's house in a village, you are not just talking to that particular member, but also the neighbours, sometimes the entire community." This notion of diffuse boundaries between a patient and their family or between a patient and their community was also strongly reflected in our data (for example in the sub-themes *patient and family as functional unit* and *community as decision-making unit*).

Therefore, a credible health chatbot that is designed for these regions will need to capture relational notions of family and community (perhaps even by collecting and sharing local stories that reinforce quality health information, as discussed later). In this way, it could have positive effects on countering mis-information in ways that reverberate outward through a community.

### 6.1.1 Providing an on-ramp and support for formal care

Based on their findings in Ecuador, Carlo et al. (2020) posit that mobile technology can bridge a gap between clinical and non-clinical care: "mobile phones can become an important artefact of the healthcare infrastructure bridging the gap between clinical and non-clinical settings. This might help support the navigation work done by patients and healthcare professionals and the interoperability across public and private infrastructures, in particular, in low-resource settings."

As such, CAI (for example via mobile phone-based messaging systems) could remove dependency on formal care in cases where health information is a sufficient intervention, but also, it can provide an on-ramp to formal care in cases where professional intervention is essential. The opportunity to provide an on-ramp and care bridge also relates to our findings around reluctance to accessing formal care. According to our data, this reluctance can stem from material barriers to accessing care (distance, cost), previous negative experiences with formal care, or a lack of knowledge about what formal care has to offer. Some of these reasons for reluctance could be addressed by CAI. For example, CAI attuned to these issues could frame interaction in ways that: a) are sympathetic to these barriers, b) can reduce some of the care gaps (e.g. by providing informational support between appointments or when formal care can't be reached) and c) provide an on-ramp to care by encouraging access, addressing anxieties, filling knowledge gaps, and preparing patients for a successful experience.

This distinction must be navigated with caution, as our findings show that participants were concerned AI would replace formal care or reduce access to it (in a context in which access is already highly constrained). Therefore, clear lines must be drawn between CAIs role in reducing care gaps and empowering communities, and ways in which CAI cannot and should not be considered a replacement to formal care.

### 6.1.2 Providing safety from stigma

Professionals in our study expected patients to be more honest with CAI than they are with human caregivers. They cited multiple reasons for patient dishonesty with a physician including stigma associated with certain conditions or behaviours, and the presence of family members during an appointment. Relatedly, Karusala and Anderson (2022) note variations in disclosure of health information based on gender: "Gender roles were a direct factor, with in-person interactions or women-only spaces sometimes being more comfortable for women." Since CAIs can be anonymous, non-gendered, and perceived as non-judgemental, they could provide information access to those uncomfortable speaking frankly in formal care environments. As stated previously, this intermediate conversational experience could then function as a segue or on-ramp to formal care or they can be sign-posted to services that address their need for privacy (e.g. women's groups, non-judgmental family planning, support groups, specialist doctors, etc.).

### 6.1.3 Leveraging multimodal information

Professionals in our study highlighted the value in being able to communicate health information with mixed modalities, including drawings, gestures, videos, stories, and other visuals. It stands to reason that such approaches address mixed literacy levels and different ways of learning and understanding. Previous research also corroborates this, both in education research (Mayer, 2005), and in digital health. For example, Carlo et al. (2020) suggest that: "Research can explore the potential use of interactive and mobile interfaces including IoT and AI-based features in low-resource settings to improve the ED waiting time and enhance nutritional prescriptions with images and personalized food-based dietary advice."

Although when it comes to CAI we tend to think of text exchanges, images and video can also be sent through technologies like WhatsApp placing CAI in a strong position to take such a multimedia approach. Indeed, the use of WhatsApp by community health workers in India (known as "ASHAs", for Accredited Social Health Activist) (Jain, 2023) demonstrates the value of this affordance. As one worker notes: "It has since become something of a best practice for ASHAs to share visually rich articles and posters via WhatsApp. These drawings or photos stay in people's minds…Instead of sending a long message, we condense the information in a single flowchart or use infographics, and it does help."   Culturally situated research on how to do this well will be important, including, for example, studies by Cajamarca et al. (2022, 2023) which inform the design of health visualisations for small screens for older adults in Chile and Ecuador.

## 6.2   A framework for Pluriversal CAI in health

In this section we discuss the value of introducing notions of pluriversality into work on culturally-aware CAI. Drawing on our findings and previous work, we present an initial framework for "pluriversal CAI in health" that we hope can support research, practice and evaluation of CAI for health in culturally diverse contexts.

### 6.2.1   The strengths and limits of existing constructs

We approached this study with the objective of identifying factors for potential cultural misalignment that designers of CAI will need to be aware of when designing for health in a Latin American context.  There are a handful of taxonomies for cultural-awareness in the context of computational linguistics (Adilazuarda et al., 2024; Hershcovich et al., 2022; Hovy & Yang, 2021). The most recent and comprehensive is the Taxonomy of Cultural Elements (TCE)(C. C. Liu et al., 2024) which incorporates previous taxonomies from computational linguistics but also draws on cultural research in sociology, anthropology and other disciplines.

As such, we opted to work with the TCE to see if we might situate our findings within it. Of course, our code tree and the TCE have different roots and approaches. The TCE aims to break down culture into constituent parts. Our research aimed to identify culturally-related influencers of health conversation (and therefore, sites of potential cultural misalignment). The former draws on concepts from theory while the latter relies on aspects of lived experience.  Furthermore, while the TCE is domain agnostic, our investigation was domain-specific to health. Although we took the position that a richer picture would come from both approaches combined, it is not surprising that we encountered some mismatch between our code tree and the cultural elements in the taxonomy.

In attempting to reconcile the two lenses, we encountered two primary tensions: 1. The targeted scope of computational linguistics which justifiably zooms in on transactional conversational exchange leaves less room for the larger systemic elements highlighted in our findings; and 2. Limits on academic definitions of culture, which are important for theory, lose meaning on the ground in a health context where multiple ecosystems are deeply entangled with culture. We elaborate on these below.

Firstly, work in the CAI space tends to focus on a model of conversational interaction that focuses on the discrete moment of exchange between a chatbot and a user. Cultural and contextual factors are therefore considered from a more contained or transactional standpoint. For example, when introducing the notion of "relationship" as an element of culture in the TCE (C. C. Liu et al., 2024), examples given refer to factors that have relatively straightforward impacts on a one-to-one conversation, such as the hierarchy of the speakers and its effect on language forms (Hovy & Yang, 2021). For example, a health chatbot designed to offer advice in Japan should know the age of the speaker in order to use the correct language form to address them appropriately. The scope is focused on elements of speaker-to-speaker interaction rather than on interdependence with relationships more broadly, including those that may not be present during the conversation but still affect it. This broader notion of relationships describes elements that were central in our findings.

For example, our data shows that the interrelated dynamics among family members affects disclosure, emotional state, and decision-making. Moreover, a chatbot may be addressing multiple family members at once, either literally—as when chatbot use is a shared experience--or figuratively, since a health decision depends, not on an isolated individual, but on a family as a decision-making unit.  A chatbot designed to support things like health behaviour change will need to take that into account.

Similarly, while the TCE includes "values" of the chatbot user as an important element, our data showed that the values of their families are also very important. As such, a health chatbot designed to offer health advice in Peru would benefit from knowing that there are multiple relational factors influencing conversation, including the beliefs of various members of a family, an individual's sense of responsibility to that family (which includes how health decisions will affect them), and how a condition may impair their ability to *support* their family or *be supported* by their family. These go well beyond the issue of variations in linguistic form and beyond an easily definable set of values.

Of course, these interrelations are not easily translated into technical goals. This reflects the tension between the complexity of the real world and the need to reduce complex entangled concepts to clean logic to advance language systems. This challenge has been acknowledged, for example, by Hovey and Yang (2021in relation to social norms: "Social norms are subtle constructs that are not easy to define, we still do not have many computational techniques to reliably quantify them, let alone assessing whether certain model behaviors should be rewarded or sanctioned."

The second tension we encountered has to do with the limits of "culture" as an academic category. Culture is often defined in ways that helpfully distinguish it from things like infrastructure, economic systems, political movements or material realities. However, our data revealed that notions of culture are so entangled with these other systems that a strict and exclusive definition loses meaning on the ground where culture and its influence emerge in ways that are inextricable from other social and material realities. For example, lack of access to medicine in pharmacies, the geographical prevalence of certain diseases like dengue, difficulty with transport in jungle regions, access to public versus private healthcare-- these factors are intertwined with culture and shape the ways people understand, talk about and enact healthcare in localised ways.  Therefore, these systems need to be considered along with elements of culture, holistically, for a coherent user experience of "cultural-appropriateness" to be achieved. In other words, if your chatbot recommends that you buy an electronic air filter for your asthma and you live in a house with no glass in the windows and you have regular power outages, it will feel like a "culturally inappropriate" suggestion (e.g. "what do you know about living in Cuba, chatbot? Clearly nothing!") even if the failure was technically related to geography or economic factors.

Therefore, we contend that, while a strict definition of culture is important for many contexts, when it comes to shaping user experience with an AI chatbot for health, a more holistic framework is necessary. Working as a team through a number of card sorting exercises and iterative discussions, we assembled a framework for such a holistic approach. It brings together the elements of the TCE with the broader themes emerging from our research.

### 6.2.2   The Framework

As mentioned above, to be culturally appropriate, a CAI will need to be contextually appropriate, where this includes cultural, social, economic, material, geographic, and logistical ecosystems. As the terms culturally aware or appropriate are unable to communicate this holism, we sought out a broader umbrella. The term "pluriversal" speaks more inclusively to this holism as it connotes worldviews, ways of knowing, and ways of experiencing the world. These certainly include culture but don't exclude other regional patterns of experience. Critically for our findings, the notion of pluriversality also incorporates a relational perspective. Therefore, we chose it to encapsulate a framework that is broad enough to meaningfully include Liu et al.'s taxonomy of cultural elements, but also the material and social infrastructures enmeshed with these that re highlighted from participatory work.

Figure 2 shows the Framework for Pluriversal Conversational AI in Health. It includes three realms of experience, each of which includes themes that influence health conversation and therefore potential sites for cultural mis-alignment. Table 3 provides more detail on each realm and includes examples of elements that constitute each theme.

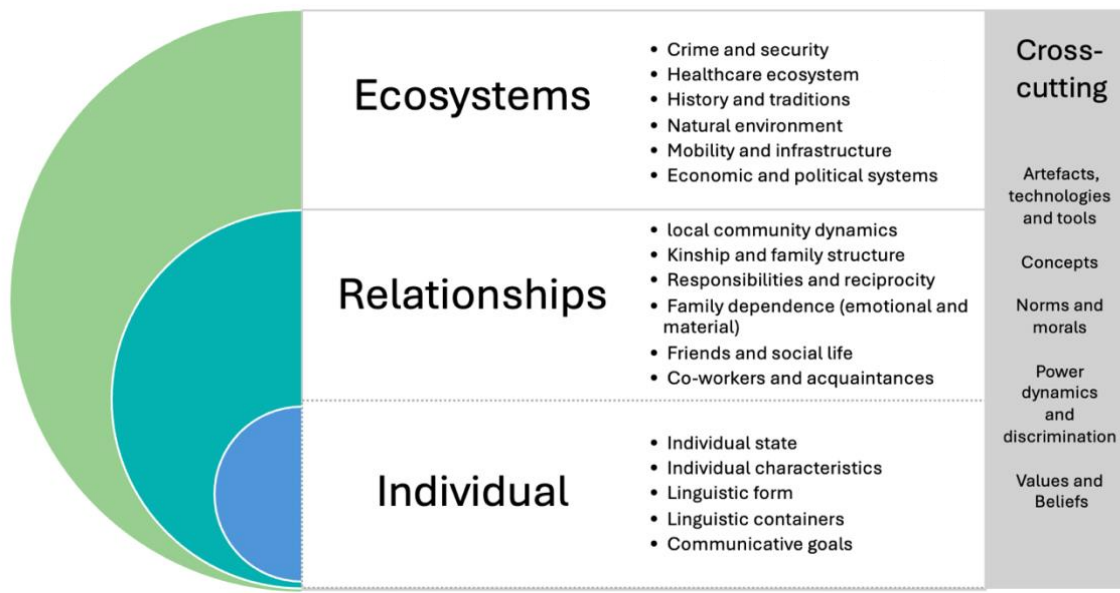

**Figure 2**. Framework for Pluriversal Conversational AI in Health. Within each of the three realms is a set of themes. Themes that cut across realms are shown at right.

The framework presents components in concentric circles to loosely distinguish between elements that refer more directly to the *individual* from elements that refer more to *relationships* and the larger *ecosystems* in which these are situated. From a practical perspective, this can help separate information about the individual (i.e. demographics, location, prior knowledge, etc.) from information about their relationships (i.e. family structure, community, responsibilities, social dependencies etc.) from information about the larger economic, political, and material systems that influence their experience of health (i.e. healthcare system, mobility, economic systems, etc.). There is also a category of cross-cutting themes that emerge across all three realms. These include four themes drawn from the TCE: 1. Artefacts, technologies and tools (we added the additional terms "technologies and tools" to "artefacts" for greater clarity), 2. concepts, 3. norms and morals, 4. values and beliefs (we added the words "and beliefs" to "values" to more clearly highlight the beliefs component of this element). We also added one additional cross-cutting theme not already in the TCE: "Power dynamics and discrimination" as this theme represents a major concern within work on culturally-appropriate CAI and we believe that it is important to give it explicit focus. These 5 themes are relevant to each of the three realms; for example, a health conversation can be influenced by an *individual's* values and beliefs, by the values and beliefs of their family (*relationships*), or by the values and beliefs arising from relevant historical traditions (*ecosystems*), even if those don't correspond with their own.

As mentioned, all elements of the TCE are incorporated into our framework. However, adaptations were made to four of the elements to adjust for our broader and health-specific context, as below.

1. "Relationship" is replaced by the top-level category "Relationships" in our framework which itself breaks into a series of important themes.
2. "Demographics" (e.g. income, gender, nationality), didn't capture our data that surfaced other individual characteristics not typically considered demographics (e.g. medical history, personality, lifestyle factors). We therefore replaced this with the theme "Individual characteristics" which includes demographics but is not limited to it.
3. "Knowledge" was expanded to "Knowledge and information needs" which considers information *preferences* as well, for example, for receiving more versus less detail about a medical issue.
4. "Context" in the TCE is defined as: "the 'containers' of communications which can be linguistic such as surrounding sentences or extra-linguistic including social settings, non-verbal cues (e.g., gesture), or historical contexts (e.g., colonization)." Based on the importance and range of extra-linguistic contexts emerging from participatory work, we felt squeezing them into one element in combination with linguistic containers obscured their importance. As such, we include the more specific theme "linguistic containers" for this element of the TCE, but the extra-linguistic components are included across multiple realms and themes in our framework.

There are, of course, countless ways these concepts could be organised, each with its own strengths and limitations. However, we present a first iteration that we ultimately found helpful in our own work for: 1. Organising the data from our study in a meaningful way that takes into account existing taxonomies, 2. Understanding the data from similar studies and comparing them with ours and, 3. moving forward with our work in culturally-appropriate language models for health by providing areas for future investigation and technological innovation.

| Realm | Theme | Example elements |
|---|---|---|
| Ecosystems | Safety and security | Safety, crime, corruption, conflict (e.g. affecting access to health resources) |
| Ecosystems | Healthcare ecosystem | Private and public systems, traditional and alternative medicine, access to medication, health providers (i.e. regional health centres, pharmacies, hospitals, etc.) |
| Ecosystems | History and traditions | Historical context, traditional ways of knowing, living and doing things |
| Ecosystems | Natural environment | Climate, food availability, regional diseases (e.g. tropical), geography |
| Ecosystems | Mobility and physical infrastructure | Transport, logistics, access to resources |
| Ecosystems | Economic and political ecosystems | Policies, regulations, funding |
| Relationships | Community dynamics | Role of community in information sharing, decision-making, etc. |
| Relationships | Co-workers and workplace | Role of workplaces and co-workers in health |
| Relationships | Family interdependence (emotional and material) | Financial dependence, practical dependence (i.e. carers), emotional support |
| | Friends and social life | Role of friends, social activities and social groups |
| | Kinship and family structure | e.g. large Intergenerational family, single parent family, same-sex couple, etc.; systems of kinship |
| | Responsibilities and reciprocity | Responsibilities to family (e.g. to care for, financially support, show reverence to, etc.) |
| Individual | Communicative goals | "the intention behind language use (e.g., requests, apologies)."* |
| | Individual state | Location (social, physical), emotional state, accessibility requirements, medical context |
| | Individual characteristics | Demographics**, lifestyle, medical history, personality (e.g. shy) |
| | Knowledge and information needs** | Education, literacy, prior knowledge, information preferences (e.g. all the details v. plain and simple) |
| | Linguistic form | "'how' to construct an utterance" * |
| | Linguistic containers | surrounding sentences** |
| Cross-cutting | Artefacts, technologies and tools | "'materialized' items as the productions of human culture, they can be forms of art, tools, machines, etc."* |
| | Concepts | "basic units of meaning underlying objects, ideas, or beliefs."* |
| | Norms and morals | "set of rules or principles that govern people's behaviour and everyday reasoning" * |
| | Power and discrimination | Bias, axes of discrimination, health disparities, inequalities, hierarchies, social classes. |

**Table 2**: Spheres, themes and example constituents of each of the themes shown in the framework diagram.

* From Liu et al.

** From Liu et al. and adapted as described in the text.

## 6.3  Relationality for pluriversal CAI

As discussed above CAI interactions are typically framed as one-to-one conversations centred on an individual user and their immediate context. However, the use of shared devices, along with larger households and more relational cultural values across many regions suggests there is value in exploring a more relational approach to CAI.

Our Latin American participants highlighted the centrality of relationships in health decision-making, health experience and within the health ecosystem. We found that the family (and sometimes the community) is a collective decision-making unit from which a patient cannot meaningfully be extracted in an isolated way. This is mirrored in Jain (2023) in which a health worker in India is quoted describing the necessity to educate, not just a pregnant woman, but also her in-laws in order to allow her to make decisions that would improve her health:

Patil had been taking photos of hundreds of regional newspaper articles addressing common health misinformation that were written by doctors… After 10 such messages, she finally had an impact; the family allowed the woman to follow her advice, and within 12 days, her hemoglobin levels had increased.

This work also shows how the influence of relationships can be harnessed for health. They describe how ASHAs share case studies of real patients who have followed their advice, such as a friend or someone they trust, showing how networks of trust lend credibility to health information.

Relatedly, Karusula and Anderson (2022) show how autonomy is relational—specifically, that trustworthy health information can be empowering for people by supporting their intentions to care for their families and communities. Their findings show that information sharing through social media increased participants' ability to take care of themselves and their family which helped reduce formal engagement with the healthcare system (which, as mentioned, can be logistically unfeasible, costly, stigmatized, or not fully trusted).

The same study also highlights that a reason people share health information is an intrinsic desire to share helpful, hopeful and trustworthy information. This suggests a CAI for health could allow for sharing of its information, especially as it establishes itself as a credible resource within a community.

Clearly, privacy preserving ways to offer people the option to share credible information and positive experiences (i.e. case studies) within their community in order to help others could be a part of a more relational CAI for health. In supporting autonomy through support for relationships and connection, a CAI has the potential to function as a *convivial* tool in Illich's terms (Illich, 1973). Such a system would allow people to take care of each other as well as themselves.

# 7   Limitations and future directions

Inevitably, any work seeking to widen representation and understanding across cultural contexts can never be comprehensive. The risk of overgeneralisation is ever present. Even so-called 'global surveys' are limited to populations of survey takers and the specific countries and regions accessed, as well as by the constraints of pre-determined closed questions. We have attempted to contribute to existing literatures by highlighting voices from several regions of Latin America, but the focus was largely on two countries, certain locations within these countries, and our engagement assumed Spanish fluency. Furthermore, our samples were constrained to those who could access the study event and effects of recruitment (e.g. for the professional workshop in Huancayo, most participants were medical students, rather than practicing medical professionals).

Therefore, while we are keenly aware results can't be said to represent the rest of the continent (or even the remainder of the countries), comparing rich results from culturally different locations does demonstrate the extent of diversity and contributes insight with respect to some of the many patterns across that diversity that are shared (we have tried to identify these shared aspects by drawing links to other studies from the majority world). Future work exploring other cultural sites in other ways (through varied methods) will continue to enrich this multidimensional picture.

Finally, our study highlights a number of open questions for computing professionals. For example: How can CAI models be developed or adapted to effectively model relational dynamics (family, community influences) in health conversations? How can models effectively integrate information like local logistics, economic constraints (e.g., availability/price of medicine), or geographic factors into their responses? How can CAI be designed to effectively leverage multimodal information for health communication in low-resource settings? Each of these frames valuable potential areas for future research.

## 7.1 Toward pluriversal CAI

A conceptual advantage of the term "pluriversal" versus "culturally-aware" is that the latter implies knowledge of a finite set of elements while the former implies co-existence among multitudes. While we may often think of cultures as being finite in number (perhaps we could train a system to understand them all?), in Peru alone, worldviews, material realities, relational dynamics and lifeworlds are multitudinous. Undoubtedly, aiming to improve CAIs' understanding of diverse cultures is essential, as the status quo is woefully imbalanced. However, individuals are culturally heterogenous, ambiguous, contradictory, and patterns of experience aren't neatly cleaved by political boundaries. As such, even if our systems were to achieve a kind of "cultural omniscience", it could never entirely solve the problem of individually tailored cultural appropriateness. So, we must ask, what alternatives are there?

Perhaps, rather than aspiring to hyper-customisation, we can learn from disciplines that have developed strategies for meaningful cross-world communication that doesn't impose values or make assumptions. While a kind of perfect neutrality is unattainable, literature and practice in intercultural communication, international relations, journalism, and psychotherapy could provide insights for CAI in highly diverse contexts, even in the absence of data about that context. Liu et al. also encourage "the integration of insights and practices from fields beyond NLP" and point to work on cultural sensitivity from clinical psychology intervention.

What would a CAI based on humility and tolerance rather than cultural mirroring be like? Perhaps it would communicate based on principles of openness and mutual respect that employ linguistic strategies for learning about their interlocutor through conversation. Like other approaches, incorporating strategies for tolerance and neutrality will have its own risks and weaknesses. However, we believe it can also contribute valuably to the collective endeavour toward more beneficial and cross-cultural human-AI interactions.

# 8 Conclusion

In this paper we have described a participatory study on conversational AI for health in Latin America. Our aim has been to contribute to understanding of the needs, preferences and opportunities for CAI interventions in health in an understudied region, as well as to identify dimensions of potential cultural misalignment in this region, with potential for implications across the majority world. Our findings revealed the need for a holistic approach to culturally-appropriate CAI that accounts for the way culture is inextricably entangled with other ecosystems in the health context. Toward this end, we present the Pluriversal Framework for CAI in Health as a way to map this holistic landscape. We hope it can support work in CAI for health and evolve as additional work across cultures reveals new dimensions and insights.

# 9 Acknowledgements

RAC and DP have been supported by the Leverhulme Centre for the Future of Intelligence via a grant from the Leverhulme Trust (grant #: RC-2015-067). RAC, MDR and FE were supported by the National Institute for Health and Care Research (NIHR) [award ID: NIHR150287].

We would also like to gratefully acknowledge everyone who supported organisation of the workshops and to all the workshop participants who generously dedicated, not only a full day of their time to participate but also shared their perspectives and lived experiences. For London workshops, we wish to acknowledge the support of the Latin American Staff community at Imperial. Dr Nervo Verdezoto Dias was instrumental in facilitating recruitment in Cardiff. In Lima, special acknowledgement is due to Lilia Cabrera for organising the community health workers, supporting recruitment efforts, and providing the space to carry out workshops. In Huancayo, we wish to acknowledge Rosario Matos Carbajal for coordinating the community health workers and facilitating participant recruitment. In Córdoba and Carlos Paz, we greatly appreciate Mariela Marchisio and Paula Etkin's invaluable support in securing the location and leveraging her networks to aid recruitment.  We also want to thank all the community health workers who contributed their precious time and personal experiences to this project.

# 10 Disclosure statement

The authors report there are no competing interests to declare.